\documentclass[aps,showpacs,superscriptaddress,twocolumn]{revtex4}%
\usepackage{amsfonts}
\usepackage{amsmath}
\usepackage{amssymb}
\usepackage{graphicx}%
\begin{document}
\title{Topological winding properties of spin edge states in Kane-Mele graphene model }
\author{Zhigang Wang}
\affiliation{LCP, Institute of Applied Physics and Computational Mathematics, P.O. Box
8009, Beijing 100088, People's Republic of China}
\author{Ningning Hao}
\affiliation{Institute of Physics, The Chinese Academy of Sciences, Beijing 100080,
People's Republic of China}
\author{Ping Zhang}
\thanks{To whom correspondence should be addressed. Email address: zhang\_ping@iapcm.ac.cn}
\affiliation{LCP, Institute of Applied Physics and Computational Mathematics, P.O. Box
8009, Beijing 100088, People's Republic of China}
\affiliation{Center for Applied Physics and Technology, Peking University, Beijing 100871,
People's Republic of China}


\begin{abstract}
We study the spin edge states in the quantum spin-Hall (QSH) effect on a
single-atomic layer graphene ribbon system with both intrinsic and Rashba
spin-orbit couplings. The Harper equation for solving the energies of the spin
edge states is derived. The results show that in the QSH phase, there are
always two pairs of gapless spin-filtered edge states in the bulk energy gap,
corresponding to two pairs of zero points of the Bloch function on the
complex-energy Riemann surface (RS). The topological aspect of the QSH phase
can be distinguished by the difference of the winding numbers of the spin edge
states with different polarized directions cross the holes of the RS, which is
equivalent to the $%
%TCIMACRO{\U{2124} }%
%BeginExpansion
\mathbb{Z}
%EndExpansion
_{2}$ topological invariance proposed by Kane and Mele [Phys. Rev. Lett.
\textbf{95}, 146802 (2005)].

\end{abstract}
\maketitle

Since the discovery of the integer quantum Hall effect (IQHE) about three
decades ago \cite{Klitzing}, topological invariants, which classify the
electronic states, have been well accepted as a powerful tool for
understanding the quantum many-body phases which have bulk energy gaps
\cite{Wen}. Historically, the pioneer work was performed by Thouless, Kohmoto,
Nightingale, and den Nijs (TKNN) \cite{TKNN}, who recognized that the IQHE can
be understood in terms of topological invariants known as Chern numbers
\cite{Chern}, which are integrals of the $k$-space Berry curvatures of the
bulk states over the magnetic Brillouin zone. While IQHE finds its elegant
connection through the adiabatic curvature with bulk topological invariants,
Halperin \cite{Halperin} first stressed that the existence of the sample
edges, which produces the current-carrying localized edge states in the Landau
energy gap, is essential in the Laughlin's gauge invariance argument
\cite{Laughlin}. Hatsugai further developed a topological theory of the edge
states \cite{Hatsugai}, in which topological invariants are the winding
numbers of the edge states on the complex-energy Riemann surface (RS).

Very recently, another striking topological quantum phenomenon, i.e., the
quantum spin Hall effect (QSHE), was identified \cite{Kane1,Kane2} after
long-distance efforts on metallic and conventionally insulating spin Hall
effects \cite{Muk1,Sinova,Mura2004}, and soon after has been attracting
extensive current interest
\cite{Onoda2005,Bern20061,Qi2006,Sheng2006,Fuk2007,Fu1,Fu2,Fu3,Mura2006,Onoda2007,Wu2006,Xu2006,Bern20062,Moore2007}
due to its basic physics and its potential application in dissipationless
spintronics. Unlike charge IQHE, whose presence fundamentally rely on the
breaking of the time-reversal ($\mathcal{T}$-) symmetry via external magnetic
field \cite{Klitzing} or intrinsic magnetic gauge flux \cite{Haldane,Ohgushi},
QSHE does not violate the $\mathcal{T}$-symmetry, which implies the absence of
the non-zero Chern invariants in QSHE insulators. Then, it turns out that a
$\mathbb{Z}_{2}$-valued topological invariant could be associated with QSHE
\cite{Kane2}. This $\mathbb{Z}_{2}$ topology is, as one selective choice,
characterized by whether the number of Kramers doublet localized at the edges
in a strip geometry is even (nonzero) or odd. If even, the insulating phase is
an ordinary Bloch insulator; otherwise, the insulating phase is a QSHE insulator.

To study QSHE and $\mathbb{Z}_{2}$ topological order, Kane and Mele introduced
the model of graphene \cite{Kane1}, which consists of two copies of Haldane's
model \cite{Haldane}, one for spin-up electrons along some axis and one for
spin-down electrons. $\mathcal{T}$-symmetry can be maintained if the intrinsic
IQHE magnetic fields are opposite for the two spin components. To make the
model more physical in that realistic mixing of the two spin components should
emerges, a Rashba spin-orbit coupling term is further included \cite{Kane2},
which now becomes the well-known Kane-Mele (KM) model. The KM model is the
simplest possible model in that it has four spin-split bands, which is the
minimum number required for the nontrivial phase to exist \cite{Moore2007}.
For this reason, the KM model has received the most attention in the other
studies \cite{Sheng2006,Fuk2007,Onoda2007,Essin2007}, in which the main focus
is on the boundary phase twist and disorder effect on QSHE.

In this paper, we give a topological study of the spin edge states and its
relation with QSHE by using the KM model. This study closely parallels with
Hatsugai's topological theory of edge-state IQHE in that we are seeking the
winding numbers of the spin edge states on the complex-energy RS. We show that
the spin-edge-state energy loops cross the holes of the RS, generating winding
numbers $I_{\uparrow}$ and $I_{\downarrow}$ for spin-up and spin-down
electrons, respectively. The quantized charge (spin) Hall conductance could be
expressed as a summation (difference) of the spin-up and spin-down winding
numbers. Thus we propose an edge-state topological invariant $I_{s}%
\mathtt{=}I_{\uparrow}\mathtt{-}I_{\downarrow}$ to distinguish a quantum
spin-Hall (QSH) insulator from an ordinary insulator. If $I_{s}$ is zero, the
insulating phase is an ordinary insulator; otherwise, the insulating phase is
a QSH insulator. We stress that this classification between topological and
ordinary insulating phases survives but is strongly modified by the weak
mixing of the two spin components, which reflects the fact that although the
exact quantization of the spin Hall conductance is destroyed by the
spin-nonconserved perturbation, the QSH phase is still topologically distinct
from the ordinary insulating phase.

Now we consider the tight-binding KM model of graphene \cite{Kane1,Kane2},
which generalizes Haldane's model \cite{Haldane} to include spin with
$\mathcal{T}$-invariant spin-orbit interactions,%
\begin{align}
H  &  =t\sum_{\langle ij\rangle\alpha}c_{i\alpha}^{\dag}c_{j\alpha}%
+i\lambda_{\text{SO}}\sum_{\langle\langle ij\rangle\rangle\alpha\beta}\nu
_{ij}c_{i\alpha}^{\dag}s_{\alpha\beta}^{z}c_{j\beta}\label{Hr}\\
&  +i\lambda_{R}\sum_{\langle ij\rangle\alpha}c_{i\alpha}^{\dag}\left(
\mathbf{s}\times\mathbf{\hat{d}}_{ij}\right)  _{z}c_{j\beta}+\lambda
_{\upsilon}\sum_{i}\xi_{i}c_{i\alpha}^{\dag}c_{i\alpha}.\nonumber
\end{align}
The first term is the nearest neighbor hopping term on the graphene
(honeycomb) lattice, while the second term is the mirror symmetric spin-orbit
interaction which involves spin-dependent second neighbor hopping. $\nu_{ij}%
$=$(2/\sqrt{3})(\mathbf{\hat{d}}_{1}\times\mathbf{\hat{d}}_{2})_{z}$=$\pm1$,
where $i$ and $j$ are next-nearest neighbors, and $\mathbf{\hat{d}}_{1}$ and
$\mathbf{\hat{d}}_{2}$ are unit vectors along the two bonds that connect $i$
to $j$. $s^{z}$ is a Pauli matrix describing the electron's spin. The third
term, which will arise due to a perpendicular electric field or interaction
with a substrate, is a nearest-neighbor Rashba coupling term, which explicitly
violates the $z\mathtt{\rightarrow-}z$ mirror symmetry. The fourth term is a
staggered sublattice potential ($\xi_{i}$=$\pm1$), which violates the symmetry
under twofold rotations in the plane.\begin{figure}[ptb]
\begin{center}
\includegraphics[width=1.0\linewidth]{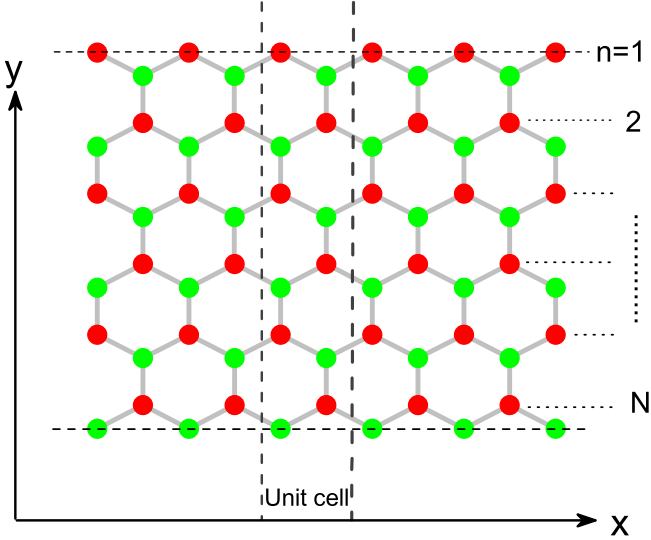}
\end{center}
\caption{(Color online) Structure of a graphene ribbon with zigzag edges,
consisting of sublattices $A$ and $B$. The width of the graphene ribbon is
$N$. Every unit cell contains $N$ numbers of $A$ and $B$ sublattices.}%
\end{figure}

For the bulk graphene, this Hamiltonian can be written in the momentum space.
For each $\mathbf{k}$ the Bloch wave function is a four-component eigenvector
$|u(\mathbf{k})\rangle$ of the Bloch Hamiltonian matrix $\mathcal{H}%
(\mathbf{k})$,%
\begin{equation}
\mathcal{H}(\mathbf{k})=\left(
\begin{array}
[c]{cccc}%
Z-\lambda_{\upsilon} & Y-iX & 0 & ia_{-}\\
Y+iX & -Z+\lambda_{\upsilon} & -ia_{+}^{\ast} & 0\\
0 & ia_{+} & -Z-\lambda_{\upsilon} & Y-iX\\
-ia_{-}^{\ast} & 0 & Y+iX & Z+\lambda_{\upsilon}%
\end{array}
\right)  \label{Hk}%
\end{equation}
with $X$=$t\sum_{i}\sin(\mathbf{k\mathtt{\cdot}a}_{i})$, $Y$=$t\sum_{i}%
\cos(\mathbf{k\mathtt{\cdot}a}_{i})$, $Z$=$-2\lambda_{\text{SO}}\sum_{i}%
\sin(\mathbf{k\mathtt{\cdot}b}_{i})$, and $a_{\pm}$=$\lambda_{R}(e^{\pm
i\frac{\pi}{3}}e^{i\mathbf{k}\cdot\mathbf{a}_{3}}\mathtt{-}e^{i\mathbf{k}%
\cdot\mathbf{a}_{2}}\mathtt{+}e^{\mp i\frac{\pi}{3}}e^{i\mathbf{k}%
\cdot\mathbf{a}_{1}})$. Here $\mathbf{a}_{1}$=$\left(  -\frac{\sqrt{3}}%
{2},-\frac{1}{2}\right)  a$, $\mathbf{a}_{2}$=$\left(  0,1\right)  a$, and
$\mathbf{a}_{3}$=$\left(  \frac{\sqrt{3}}{2},-\frac{1}{2}\right)  a$ represent
the vectors from sublattice site $A$ (red circles in Fig. 1) to its three
nearest sublattice sites $B$ (green circles in Fig. 1), respectively, and
$\mathbf{b}_{1}\mathtt{=}\mathbf{a}_{2}\mathtt{-}\mathbf{a}_{3}$,
$\mathbf{b}_{2}\mathtt{=}\mathbf{a}_{3}\mathtt{-}\mathbf{a}_{1}$, and
$\mathbf{b}_{3}\mathtt{=}\mathbf{a}_{1}\mathtt{-}\mathbf{a}_{2}$ represent the
vectors between the nearest same sublattice sites. No doubt, the Hamiltonian
(\ref{Hk}) can also be written in terms of the SO(5) Clifford algebra, which
is a very helpful technique to deduce $\mathbb{Z}_{2}$-valued topological
invariants \cite{SCZhang}.

To explore the edge topological invariant characterizing the QSH phase, we now
turn to study the graphene ribbon with zigzag edges (see Fig. 1), which is
periodic in the $x$ direction while it has two edges in the $y$ direction. In
the following, we replace index $i$ with $(mns)$ to denote the lattice sites,
where $(mn)$ label the unit cells and $s$ label the sublattice $A$ and $B$ in
this cell. Since the graphene ribbon is periodic in the $x$ direction, we can
use a momentum representation of the electron operator%
\begin{equation}
c_{mns,\alpha}=\frac{1}{\sqrt{L_{x}}}\sum_{k}e^{ikX_{(mns)}}\gamma_{ns,\alpha
}(k), \label{c}%
\end{equation}
where $(X_{(mns)},Y_{(mns)})$ represents the coordinate of the site $s$ in the
unit cell $(mn)$, and $k$ is the momentum along the $x$ direction. For
simplification, let us first consider the spin-conserved case, i.e., the
Rashba term in Hamiltonian (\ref{Hr}) is set to be zero. Also, we let the
staggered sublattice potential vanishes and only consider the first two terms
in the Hamiltonian (\ref{Hr}). Now let us consider the one-particle state
$|\psi(k)\rangle$=$\sum_{n,s,\alpha}\psi_{ns\alpha}(k)\gamma_{ns,\alpha}%
^{\dag}|0\rangle$. Inserting it into the Schr\"{o}dinger equation
$H|\psi\rangle$=$\epsilon|\psi\rangle$, one can get the following eigenvalue
equations for $A$- and $B$-sublattice sites:%
\begin{align}
(\epsilon+p_{2})\psi_{nA\uparrow}  &  =p_{1}\psi_{nB\uparrow}+t\psi
_{(n-1)B\uparrow}\nonumber\\
&  +p_{3}\left[  \psi_{(n+1)A\uparrow}+\psi_{(n-1)A\uparrow}\right]
,\label{1(a)}\\
(\epsilon-p_{2})\psi_{nA\downarrow}  &  =p_{1}\Psi_{nB\downarrow}%
+t\Psi_{(n-1)B\downarrow}\nonumber\\
&  -p_{3}\left[  \Psi_{(n+1)A\downarrow}+\Psi_{(n-1)A\downarrow}\right]
,\label{1(b)}\\
(\epsilon-p_{2})\psi_{nB\uparrow}  &  =p_{1}\psi_{nA\uparrow}+t\psi
_{(n+1)A\uparrow}\nonumber\\
&  -p_{3}\left[  \psi_{(n+1)B\uparrow}+\psi_{(n-1)B\uparrow}\right]
,\label{1(c)}\\
(\epsilon+p_{2})\psi_{nB\downarrow}  &  =p_{1}\psi_{nA\downarrow}%
+t\psi_{(n-1)A\downarrow}\nonumber\\
&  +p_{3}\left[  \psi_{(n+1)B\downarrow}+\psi_{(n-1)B\downarrow}\right]  ,
\label{1(d)}%
\end{align}
where $p_{1}$=$2t\cos\left(  \frac{\sqrt{3}}{2}ka\right)  $, $p_{2}$%
=$2\lambda_{\text{SO}}\sin\left(  \sqrt{3}ka\right)  $, and $p_{3}%
=2\lambda_{\text{SO}}\sin\left(  \frac{\sqrt{3}}{2}ka\right)  $. Eliminating
the $B$($A$)-sublattice sites, we obtain the difference equation for $A$%
($B$)-sublattice sites
\begin{equation}
f_{1}\psi_{n}=f_{2}\left[  \psi_{n+2}+\psi_{n-2}\right]  +f_{3}\left[
\psi_{n+1}+\psi_{n-1}\right]  , \label{diff}%
\end{equation}
where $f_{1}$=$\epsilon^{2}-p_{2}^{2}-p_{1}^{2}-t^{2}-2p_{3}^{2}$, $f_{2}%
$=$p_{3}^{2}$, $f_{3}$=$p_{1}t-2p_{2}p_{3}$ and $\psi_{nA\alpha}$%
($\psi_{nB\alpha}$) was replaced by $\psi_{n}$. Eq. (\ref{diff}) is the
so-called Harper equation \cite{Harper}. Note that since the spin $s_{z}$ is
conserved, the spin-up and spin-down electrons satisfy the same Harper
equation (\ref{diff}). \begin{figure}[ptb]
\begin{center}
\includegraphics[width=1.0\linewidth]{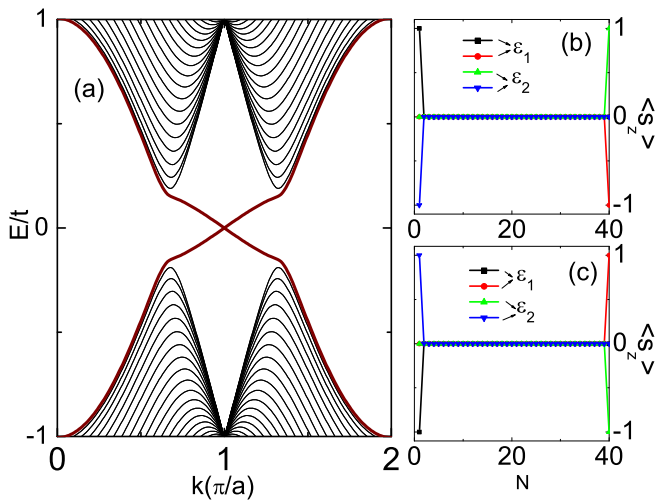}
\end{center}
\caption{(Color online) (a) Energy spectrum for the graphene ribbon with
zigzag edges. The spin-orbit coupling parameter in Eq. (\ref{Hr}) is chosen as
$\lambda_{\text{SO}}$=$0.03t$ and the last two terms are neglected. The bands
crossing the gap are spin filtered edge states, which are depicted in (b) and
(c) by plotting their spin ($s_{z}$) distribution on the lattice sites with
$k$=$0.99\pi$ and $k$=$1.01\pi$, respectively. To be more clear, the energies
used in the right panels are denoted by $\epsilon_{1}$ and $\epsilon_{2}$ for
lower and upper spin-degenerate edge states, respectively. }%
\label{2}%
\end{figure}

Now we represent Eq. (\ref{diff}) in the transfer matrix form. For this
purpose we introduce a new wave function $\varphi_{n}$, which is a linear
transformation of the original wave function $\psi_{n}$,%
\begin{equation}
\varphi_{n}=\psi_{n}\pm t_{\mp}\psi_{n-1}+\psi_{n-2}, \label{fan}%
\end{equation}
where $t_{\pm}$=$(\sqrt{b^{2}+4(2+d)}\pm b)/2$ with $b$=$-f_{3}/f_{2}$ and
$d$=$f_{1}/f_{2}$. The new wave function $\varphi_{n}$ can be written in the
following transfer matrix form%
\[
\left(
\begin{array}
[c]{c}%
\varphi_{n}\\
\varphi_{n-1}%
\end{array}
\right)  =\tilde{M}(\epsilon)\left(
\begin{array}
[c]{c}%
\varphi_{n-1}\\
\varphi_{n-2}%
\end{array}
\right)  ,
\]
with%
\begin{equation}
\tilde{M}(\epsilon)=\left(
\begin{array}
[c]{cc}%
\pm t_{\pm} & -1\\
1 & 0
\end{array}
\right)  . \label{m1}%
\end{equation}
Although $\varphi_{n}$ is a transformation of the original wave function
$\psi_{n}$, here we also can take $\varphi_{0}$ and $\varphi_{N}$ as the wave
functions of two edges, which are linked by a reduced transfer matrix as
follows:%
\begin{equation}
\left(
\begin{array}
[c]{c}%
\varphi_{N+1}\\
\varphi_{N}%
\end{array}
\right)  =M(\epsilon)\left(
\begin{array}
[c]{c}%
\varphi_{1}\\
\varphi_{0}%
\end{array}
\right)  , \label{tm}%
\end{equation}
where%
\begin{equation}
M(\epsilon)=\left[  \tilde{M}(\epsilon)\right]  ^{N}=\left(
\begin{array}
[c]{cc}%
M_{11} & M_{12}\\
M_{21} & M_{22}%
\end{array}
\right)  . \label{M}%
\end{equation}
All kinds of solutions of Eq. (\ref{M}) are obtained by different choices of
$\varphi_{0}$ and $\varphi_{1}$. Now we study the energy spectrum of the
graphene ribbon with special attention to the edge states. The general
boundary condition is
\begin{equation}
\varphi_{N+1}=\varphi_{0}=0. \label{bc}%
\end{equation}
With Eqs. (\ref{tm}), (\ref{M}) and the boundary condition (\ref{bc}), one can
easily get that the solutions satisfy%
\begin{equation}
M_{21}(\epsilon)=0, \label{m21}%
\end{equation}
and
\begin{equation}
\varphi_{N}=M_{11}(\epsilon)\varphi_{1}. \label{m11}%
\end{equation}
If we use a usual normalized wave function, the state is localized at the
edges as%
\begin{equation}
\left\{
\begin{array}
[c]{cc}%
|M_{11}(\epsilon)|\ll1 & \text{localized at }n\approx1\text{ (down edge),}\\
|M_{11}(\epsilon)|\gg1 & \text{localized at }n\approx N\text{ (up edge).}%
\end{array}
\right.  \label{du}%
\end{equation}

Because the analytical derivation is very difficult, we now turn to start a
numerical calculation from Eq. (\ref{Hr}) with $\lambda_{R}$ and
$\lambda_{\upsilon}$ setting to be zero. Figure 2(a) shows the energy bands
for the graphene with zigzag edges with the intrinsic spin-orbit coupling
strength $\lambda_{\text{SO}}$=$0.03t$, same as that in Ref. \cite{Kane1}.
From Fig. 2(a) one can see that there are two branches of spin-degenerate
dispersed energy bands with two pairs of spin states lying in the energy gap.
These two pairs of spin states cross at the $\mathcal{T}$-invariant point
$k\mathbf{=}\pi$ and are localized at the edges of the graphene ribbon. To
show the spin localization features of these gapless spin states, we plot in
Figs. 2(b) and 2(c) the spin ($z$-component) distribution along the graphene
strip for the two pairs of edge states at $k\mathtt{=}$0.99$\pi$ and 1.01$\pi
$, respectively. At the left side of the crossing point $k\mathbf{=}\pi$ [Fig.
2(b)], the spin-up (-down) edge state with lower energy $\epsilon_{1}$ is
localized at the down (up) edge of the graphene ribbon, while the spin-up
(-down) edge state with upper energy $\epsilon_{2}$ is localized at the up
(down) edge of the graphene ribbon. On the other hand, at the right side of
the crossing point $k\mathbf{=}\pi$ [Fig. 2(c)], the spin-up (-down) edge
state with lower energy $\epsilon_{1}$ is localized at the up (down) edge of
the graphene ribbon, while the spin-up (-down) edge state with upper energy
$\epsilon_{2}$ is localized at the down (up) edge of the graphene ribbon.
Thus, if we trace the spin-up flow of the edge states when varying the
momentum $k$ in one period, it can be found that the electrons with different
spin propagate in opposite directions and that the electronic state of
graphene is a QSH state. The spin Hall conductivity has been shown as
$\sigma_{S}$=$2$ in units of $e/4\pi$ by different methods.

Before introducing the winding number of the spin edge states on the complex
energy RS, let us try to interpret the above two numerical characteristics.
The first one is that there are always two energy-degenerate edge states
appearing localized at opposite edges. To explain this issue, we suppose the
spin edge states to be exponentially localized on the boundary with the
following ansatz \cite{konig, creutz}
\begin{equation}
\psi_{n}=\lambda^{n}\psi, \label{ansatz}%
\end{equation}
where $\lambda$ is a complex number. Inserting Eq. (\ref{ansatz}) into the
Harper equation (\ref{diff}), one can easily get the complex number $\lambda$
satisfying the following equation%
\begin{equation}
\left(  \lambda+\lambda^{-1}\right)  ^{2}-b\left(  \lambda+\lambda
^{-1}\right)  -\left(  d+2\right)  =0. \label{lmd}%
\end{equation}
That means
\begin{equation}
\lambda+\lambda^{-1}=\pm t_{\pm}. \label{lmd1}%
\end{equation}
It is trivial to see from Eqs. (\ref{lmd}) and (\ref{lmd1}) that if $\lambda$
is a solution, then $\lambda^{-1}$ is also to be. That means if there is an
edge state localized near one boundary, at the same time there is another edge
state localized near opposite boundary. Moreover, note that the conservation
of the spin $s_{z}$, i.e., the spin-up and -down electrons satisfy the same
Harper equation (\ref{diff}), one easily find that if $\lambda$ is a solution
describing an edge state with spin-up localized near one boundary, then
$\lambda^{-1}$ is another solution describing an edge state with spin-down
localized near opposite boundary. That is the intrinsic reason for the second
feature that there are always two edge states with opposite spins localized
near opposite edges.

Now we show that similar to charge IQHE, QSHE of the graphene ribbon is
associated with winding number of the spin edge states on a complex energy RS.
Within the topological edge theory \cite{Hatsugai}, this number is, as well as
the Chern number, a well-defined topological quantity. In order to well
understand the winding numbers of the spin edge states in QSHE, let us first
consider that of the edge states in charge IQHE. First we ignore the open
condition and consider the bulk Bloch function at sites with $y$-coordinate of
$L_{y}$. For Bloch function, $\mathbf{\varphi}_{0}^{(b)}$\ and
$\mathbf{\varphi}_{1}^{(b)}$\ compose an eigenvector of $M$\ with the
eigenvalue $\rho$,%

\begin{equation}
M(\epsilon)\left(
\begin{array}
[c]{c}%
\mathbf{\varphi}_{1}^{(b)}\\
\mathbf{\varphi}_{0}^{(b)}%
\end{array}
\right)  =\rho\left(  \epsilon\right)  \left(
\begin{array}
[c]{c}%
\mathbf{\varphi}_{1}^{(b)}\\
\mathbf{\varphi}_{0}^{(b)}%
\end{array}
\right)  . \label{e16}%
\end{equation}
In order to discuss the wave function of the edge state, we extend the energy
to a complex energy. In the following, we use a complex variable $z$\ instead
of real energy$\ \epsilon$. From Eq. (\ref{e16}) we get%

\begin{equation}
\rho(z)=\frac{1}{2}\left[  \Delta(z)-\sqrt{\Delta^{2}(z)-4}\right]
\label{e17}%
\end{equation}
and%

\begin{equation}
\mathbf{\varphi}_{N}=-\frac{M_{11}\left(  z\right)  +M_{22}(z)-\omega}%
{-M_{11}\left(  z\right)  +M_{22}(z)+\omega}M_{21}(z)\mathbf{\varphi}_{1},
\label{e18}%
\end{equation}
where $\Delta(z)$=Tr$\left[  M(z)\right]  $ and $\omega$=$\sqrt{\Delta
^{2}(z)-4}$. Clearly,%

\begin{equation}
\det M(\epsilon)=1 \label{e19}%
\end{equation}
since $\det\tilde{M}(\epsilon)$=$1$. From Eq. (\ref{e18}), one can find that
the analytic structure of the wave function is determined by the algebraic
function $\omega$=$\sqrt{\Delta^{2}(z)-4}$. The RS of $\omega$=$\sqrt
{\Delta^{2}(z)-4}$\ on the complex energy plane can be built by the
conglutination between different analytic brunches. The close complex energy
plane can be obtained from the open complex\ energy plane through spheral pole
mapping. Now let us discuss the analytic structure of $\omega=\sqrt{\Delta
^{2}(z)-4}$ on the open complex plane. If the system has $q$\ energy bands, i.e.,%

\begin{equation}
\epsilon\in\left[  \lambda_{1},\lambda_{2}\right]  ,...,\left[  \lambda
_{2j-1},\lambda_{2j}\right]  ,...,\left[  \lambda_{2q-1},\lambda_{2q}\right]
, \label{e20}%
\end{equation}
where $\lambda_{j}$\ denote energies of the band edges and $\lambda
_{i}<\lambda_{j}$, $i<j$, then $\omega$\ can be factorized as%

\begin{equation}
\omega=\sqrt{\Delta^{2}(z)-4}=\sqrt{%
%TCIMACRO{\dprod \limits_{j=1}^{2q}}%
%BeginExpansion
{\displaystyle\prod\limits_{j=1}^{2q}}
%EndExpansion
\left(  z-\lambda_{j}\right)  }. \label{q1}%
\end{equation}
In the present graphene system, there are two energy bands, so $q$=$2$. The
two single-valued analytic branches are defined on the plane with two secants
down or up the bank of which we choose corresponding complex angles, which is
shown in Fig. 3.

\begin{figure}[ptb]
\begin{center}
\includegraphics[width=0.8\linewidth]{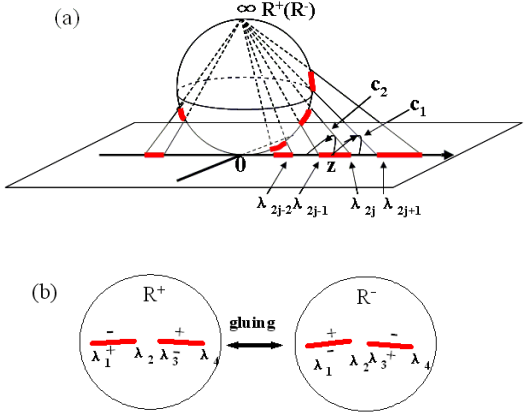}
\end{center}
\caption{(Color online) (a) The open complex energy plane are mapped to the
close complex energy plane through spheral pole projection. (b) Two sheets
with two cuts which correspond to the energy bands of the graphene
nanoribbons. The RS of the Bloch function is obtained by gluing the two
spheres along the arrows near the cuts. }%
\end{figure}

In order to ensure $\omega(\mu_{j})\geq0$ ($j$=$1$,$...$,$g$=$q-1$) [Here,
$\mu_{j}$ is the energy of the edge state in the gap $\left[  \lambda
_{2j},\lambda_{2j+1}\right]  $], we divide the two single-value analytic
branches based on the parity of $j$. Let us consider the case that the energy
$z$ lies in the band $\left[  \lambda_{2j-1},\lambda_{2j}\right]  $ [see Fig.
3(a)]. The left side of this energy band is the\ $(j-1)$th\ gap, and the right
side is the $j$th gap. When the energy $z$\ moves from the $j$th band to the
$(j-1)$th gap (the $j$th gap) along an arbitrary path $c_{1}$ ($c_{2}$), only
the singularities $\lambda_{2j-1}$ and $\lambda_{2j}$\ have contributions to
the variance of the principal value of argument of $\omega$. On the up bank of
the secant, we distinguish two branches $R^{+}$\ and $R^{-}$ as the following:

For $j$ is even, if we set $\arg(z-\lambda_{2j-1})$=$0$ and $\arg
(z-\lambda_{2j})$=$\pi$, which corresponds to $\omega(\mu_{j-1})>0$
($\omega(\mu_{j})<0$) when $z$ moves along $c_{1}$ ($c_{2}$), then the branch
$R^{+}$ is defined as a complex plane with $q$ secants. If we set
$\arg(z-\lambda_{2j-1})$=$2\pi$ and $\arg(z-\lambda_{2j})$=$\pi$, which
corresponds to $\omega(\mu_{j-1})<0$ ($\omega(\mu_{j})>0$) when $z$ moves
along $c_{1}$ ($c_{2}$), then the branch $R^{-}$ is defined as a complex plane
with $q$ secants. For $j$ is odd, the definitions of $R^{+}$ and $R^{-}$ are
reverse to those for $j$ is even. So, if $z$\ lies in the $j$th gap from the
lowest one on the real axis,%

\begin{equation}
\alpha\left(  -1\right)  ^{j}\omega\geq0,\ z\ (\text{real})\text{ on
}R^{\alpha}\ (\alpha=+,-), \label{q2}%
\end{equation}
and at energies $\mu_{j}$ of the edge states we have%

\begin{equation}
\omega(\mu_{j})=\alpha\left(  -1\right)  ^{j}\left\vert M_{11}\left(  \mu
_{j}\right)  -M_{22}\left(  \mu_{j}\right)  \right\vert , \label{e25}%
\end{equation}
where $\mu_{j}\in R^{\alpha}$, $\alpha$=$+,-$. In addition, one can easily obtain%

\begin{equation}
\Delta(\epsilon)\left\{
\begin{array}
[c]{c}%
\leq-\text{2 for }j\text{ odd}\\
\geq\text{2 for }j\text{ even}%
\end{array}
\right.  , \label{e26}%
\end{equation}
where the energy $\epsilon$\ (on $R^{\alpha}$)\ is in the $j$th\ gap.

When the branches $R^{+}$and $R^{-}$\ on the open complex energy plane are
mapped to the close complex energy plane through spheral-pole-projection, one
can get two single-value analytic spherical surfaces. The RS is obtained by
gluing the two spherical surfaces at these branch cuts. Make sure the $\pm$
banks face the $\mp$ banks of other sphere. (see Figs. 3(b)). Note that there
are two real axes after gluing. In our model, the genus of the RS is $g$=$1$
for spin-up (or spin-down) electrons, which is the number of energy gaps. The
wave function is defined on the $g$=$1$ RS $\Sigma_{g=1}(k_{x})$. The branch
of the Bloch function is specified as $\omega>0$, which we have discussed
above. With Eqs. (\ref{e26}), (\ref{e25}) and (\ref{e18}), and using the fact
when $\mathbf{\varphi}_{L_{y}-1}(\mu_{j})=0$\ for $\mu_{j}\in R^{\alpha}$ and
$\mathbf{\varphi}_{L_{y}-1}(\mu_{j})\neq0$\ for $\mu_{j}\in R^{-\alpha}$, one
can obtain that when the zero point is on the upper sheet of RS ($R^{+}$), the
edge state is localized at the down edge; when the zero point is on the lower
sheet of RS ($R^{-}$), the edge state is localized at the up edge.

In Fig. 4, on the RS $\Sigma_{g=1}(k_{x})$ of our system the energy gap
corresponds to the loop around the hole of the $\Sigma_{g=1}(k_{x})$ and the
energy bands correspond to the closed paths vertical to the energy gap loop on
the $\Sigma_{g=1}(k_{x})$. The Bloch function is defined on this surface. For
the fixed $k_{x}$ there is always $g$=$1$\ zero point at the down-edge-state
energy $\mu_{j}$. Since there are two real axes on the $\Sigma_{g=1}(k_{x})$,
correspondingly, there is $g$=$1$\ zero point at the up-edge-state energy
$\mu_{j}$.\begin{figure}[ptb]
\begin{center}
\includegraphics[width=0.8\linewidth]{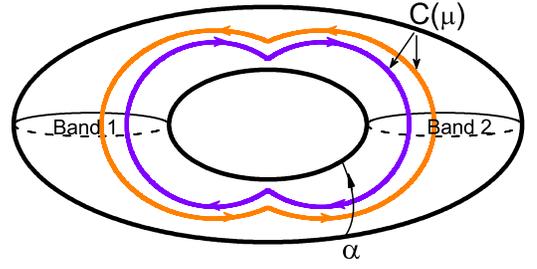}
\end{center}
\caption{(Color online) RS of the Bloch function corresponding to Fig. 2(a).
To clearly see the degenerate edge states with different spin, we
schematically separate their traces with different colors. The orange and
purple curves correspond to spin-up and spin-down channels, respectively. }%
\end{figure}

The above considerations are for the fixed $k_{x}$. Now, let us consider a
family of the RSs $\Sigma_{g=1}(k_{x})$ parametrized by $k_{x}$ changing in
one of its periods. $\Sigma_{g=1}(k_{x})$ can be modified by this change.
However, all the RSs $\Sigma_{g=1}(k_{x})$ with different $k_{x}\ $are
topologically equivalent if there are stable energy gaps in the 2D energy spectrum.

On the genus $g$=$1$\ RS,\ the first homotopy group is generated by $2g$%
=$2$\ generators, $\alpha_{i}$\ and $\beta_{i}$, $i$=$1$. The intersection
number of the curves (including directions) \cite{Hatsugai} is given by (see
Fig. 5)%

\begin{equation}
I\left(  \alpha_{i},\beta_{j}\right)  =\delta_{ij}. \label{e27}%
\end{equation}
Any curves on the RS are spanned homotopically by $\alpha_{i}$\ and $\beta
_{i}$. When $\mu_{j}(k_{x})$\ \ moves $p$\ times around the $j$th\ hole with
some integer $t$, one has%

\begin{equation}
C(\mu_{j})\approx\beta_{j}^{p}, \label{e28}%
\end{equation}
which means%

\begin{equation}
I\left(  \alpha_{i},C(\mu_{j})\right)  =t\delta_{ij}. \label{e29}%
\end{equation}
When the Fermi energy $\epsilon_{F}$\ of the 2D system lies in the $i$th
energy gap, the charge Hall conductance is given by the winding number of the
edge state, which is given by the number of intersections $I(\alpha_{j}%
,C(\mu_{j}))$ ($\equiv I(C(\mu_{j}))$)\ between the canonical loop $\alpha
_{j}$\ on the RS and the trace of $\mu_{j}$, i.e.,%

\begin{equation}
\sigma_{xy}^{j,\text{edge}}=-\frac{e^{2}}{h}I(C(\mu_{j})). \label{e30}%
\end{equation}

Similarly, the above expression can be obtained by the Byers-Yang \cite{Yang}
and Laughlin-Halperin \cite{Laughlin,Halperin} gauge arguments. The system
with the periodic boundary condition in the $x$ direction and open condition
in the $y$ direction can be considered as a cylinder. By Laughlin gauge
invariance argument, the vector potential $A$\ has to have the form
$A$=$n\frac{hc}{eL_{x}}$($n$ is an integer). When the flux $\Phi$\ threading
the cylinder is adiabatically turn on from $\Phi\mathbf{(}0\mathbf{)}%
$\textbf{=}$0$\ to $\Phi\mathbf{(}T\mathbf{)}$\textbf{=}\ $hc/e$\ with
$\Delta\Phi$\textbf{=}$hc/e$\ a flux quantum, $\Delta A$=$\frac{hc}{eL_{x}}%
$\ accords with the gauge argument, therefore, $\Delta A$($\Delta\Phi$)\ maps
the system back to itself. Basing on the single-electron assumption, When the
Fermi energy lies in the $j$th energy gap, there are $I(C(\mu_{j}))$ states
(electrons) transferring from the down edge ($y$=$1$) to the up edge
($y$=$L_{y}-1$) in net. The energy change during the adiabatic process is
$\Delta E$=$I(C(\mu_{j}))(-e)V$, where $V_{y}$\ is a voltage in the $y$
direction. This gives the charge Hall current $I_{x}$\ as follows%

\[
I_{x}=c\frac{\Delta E}{\Delta\mathbf{\Phi}}=\sigma_{xy}V_{y}.
\]
Then we get an expression for $\sigma_{xy}^{\text{edge}}$\ as Eq. (\ref{e30}).

The above analysis is for the winding number of the edge states in IQHE and it
can be easily generalized to the winding number of the spin edge states in
QSHE when the spin degree of freedom is considered. In the latter case, the
winding number of the spin edge state is given by the number of intersections
$I_{S}(\alpha_{j},C(\mu_{j}))$ ($\equiv I_{S}(C(\mu_{j}))$, $S=\uparrow$,
$\downarrow$ for spin-up and -down, respectively)\ between the canonical loop
$\alpha_{j}$\ on the RS and the trace of $\mu_{j}$. From Fig. 4 one can
observe that $\mu$\ moves one time across the hole ($j$=$1$), which means
$C(\mu)$\ $\approx\beta$\ and $\left\vert I_{S}(C(\mu))\right\vert $=$1$.
Considering the winding direction (see Fig. 5), one obtain $I_{\uparrow}%
(C(\mu))$=$1$ for spin-up electrons\ while $I_{\downarrow}(C(\mu))$=$-1$ for
spin-down ones. Thus, the charge Hall conductance is given by the summation
over two spin channels,
\begin{equation}
\sigma_{xy}^{(c)\text{edge}}=(I_{\uparrow}+I_{\downarrow})\frac{e^{2}}%
{h}\equiv I_{c}\frac{e^{2}}{h}=0, \label{qhc}%
\end{equation}
while the spin Hall conductance is given by the difference between them,
\begin{equation}
\sigma_{xy}^{(s)\text{edge}}=(I_{\uparrow}-I_{\downarrow})\frac{\hbar}{2}%
\frac{e^{2}}{h}\equiv I_{s}\frac{e}{4\pi}=2\frac{e}{4\pi}. \label{shc}%
\end{equation}

\begin{figure}[ptb]
\begin{center}
\includegraphics[width=0.8\linewidth]{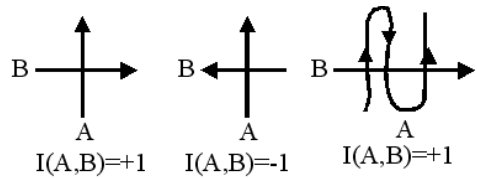}
\end{center}
\caption{Intersection number $I(A,B)$ of two curves A and B. Each intersection
point contributes by $+1$ or $-1$ according to the direction.}%
\end{figure}\begin{figure}[ptbptb]
\begin{center}
\includegraphics[width=0.7\linewidth]{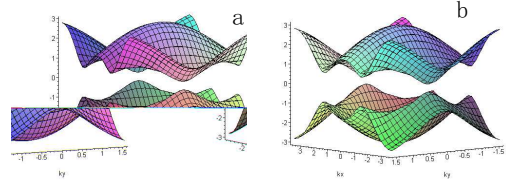}
\end{center}
\caption{(Color online) Energy spectra of the bulk\ graphene (a) in the QSH
phase with $\lambda_{v}$=$0.1t$ and $\lambda_{\text{SO}}$=$0.05t$ and (b) in
the ordinary insulator phase with $\lambda_{v}$=$0.4t$ and $\lambda
_{\text{SO}}$=$0.05t$. There are bulk gaps appearing in both two different
phases.}%
\end{figure}

When the finite on-site energy $\lambda_{\upsilon}$ is considered while the
Rashba coupling is kept as zero, the two degenerate energy bands will split.
For the bulk graphene $\mathcal{H}(\mathbf{k})$ (\ref{Hk}), one can easily
obtain the four dispersion energy bands,%
\begin{equation}
\epsilon_{\mathbf{k}}=\pm\sqrt{X^{2}+Y^{2}+(Z\pm\lambda_{v})^{2}}, \label{e}%
\end{equation}
between which there is an energy gap with magnitude $\Delta$=$2|\lambda
_{\upsilon}-3\sqrt{3}\lambda_{\text{SO}}|$. We plot in Fig. 6 the energy
dispersion with different $\lambda_{\upsilon}$ and $\lambda_{\text{SO}}$. The
condition $\lambda_{\upsilon}\neq3\sqrt{3}\lambda_{\text{SO}}$ provides a
finite bulk energy gap in the graphene. However, when the Fermi energy lies in
the gap, the graphene's phases are different with different topology: for
$\lambda_{\upsilon}>3\sqrt{3}\lambda_{\text{SO}}$, the system is an ordinary
insulator and for $\lambda_{\upsilon}<3\sqrt{3}\lambda_{\text{SO}}$, the
system is a topological insulator with QSH phase. Kane and Mele \cite{Kane2}
have proposed a $%
%TCIMACRO{\U{2124} }%
%BeginExpansion
\mathbb{Z}
%EndExpansion
_{2}$ index describing the QSH phase, which is determined by counting the
number of pairs complex zeros of $P(\mathbf{k})$=Pf$\left[  \langle
u_{i}(\mathbf{k})|\Theta|u_{j}(\mathbf{k})\rangle\right]  $ with
$|u_{i}(\mathbf{k})\rangle$ the wave functions corresponding to the bulk
Hamiltonian (\ref{Hk}) and $\Theta$ time reversal operator. To understand
these different phases, Kane and Mele also studied the edge states of the
graphene ribbon with zigzag edges. However, the winding properties of the edge
states in the KM model has not been previously considered, which just is the
special focus of our present study.

\begin{figure}[ptb]
\begin{center}
\includegraphics[width=0.6\linewidth]{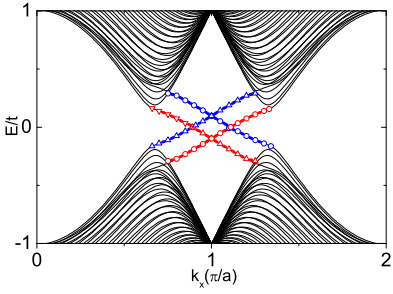}
\end{center}
\caption{(Color online) Energy spectrum of the \textquotedblleft
zigzag\textquotedblright\ graphene ribbon in the QSH phase with $\lambda_{v}%
$=$0.1t$ and $\lambda_{\text{SO}}$=$0.05t$. The red and blue lines represent
the edge states localized at the down and up edges of the system,
respectively. And the circle and triangle label the up and down spins,
respectively.}%
\end{figure}\begin{figure}[ptbptb]
\begin{center}
\includegraphics[width=0.6\linewidth]{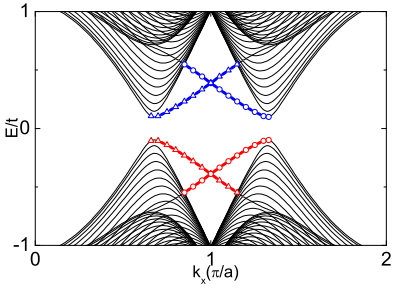}
\end{center}
\caption{(Color online) Energy spectrum of the \textquotedblleft
zigzag\textquotedblright\ graphene ribbon in the ordinary insulator phase with
$\lambda_{v}$=$0.4t$ and $\lambda_{\text{SO}}$=$0.05t$. The red and blue lines
represent the edge states localized at the down and up edges of the system,
respectively. And the circle and triangle label the up and down spins,
respectively.}%
\end{figure}

Considering nonzero $\lambda_{\upsilon}$, now let us investigate the
topological winding numbers with zigzag edges. In this case, the Harper
equation for site $A$ turns to have the same form as Eq. (\ref{diff}),
provided that the $p_{2}$ in the coefficients $f_{1}$ and $f_{3}$ is replaced
by $p_{2}\pm\lambda_{\upsilon}$ for spin-up and -down channels, respectively.
Similarly, one can introduce a new wave function, as a linear transformation
of the original wave functions (\ref{fan}), satisfying the matrix form with
the transform matrix $\tilde{M}(\epsilon)$ (\ref{m1}). The $p_{2}$ in the
coefficients $b$, $d$, and $t_{\pm}$ is now also replaced by $p_{2}\pm
\lambda_{\upsilon}$ for different spins. In evidence, the edge-state energies
become non-degenerate for the finite on-site energy $\lambda_{\upsilon}$.
Figure 7 shows the energy spectrum of the \textquotedblleft
zigzag\textquotedblright\ graphene ribbon in a QSH phase with $\lambda
_{\upsilon}$=$0.1t$ and $\lambda_{\text{SO}}$=$0.05t$, which satisfy
$\lambda_{\upsilon}<3\sqrt{3}\lambda_{\text{SO}}$. From Fig. 7, one can
clearly see that the gapless edge states are localized near the system
boundary with different spins. The corresponding complex energy RS of the
Bloch wave functions is topologically equivalent to Fig. 3. And no doubt this
QSH phase can be described by the difference of the winding numbers of the
edge states with different spins, i.e., $I_{s}$=$I_{\uparrow}-I_{\downarrow}%
$=$2$, in units of $e/4\pi$. On the other hand, one can obtain the charge Hall
conductivity $\sigma_{c}$ is zero since $I_{\uparrow}$=$-I_{\downarrow}$. As a
comparison, we draw in Fig. 8 the energy spectrum of an ordinary insulator
phase with $\lambda_{\upsilon}$=$0.4t$ and $\lambda_{\text{SO}}$=$0.05t$. One
can find that in this case (see Fig. 9), there are no gapless edge states
connecting the two energy bands. The traces of the edge states in the
corresponding complex energy RS are sunk in the bulk bands, i.e.,
$I_{\uparrow}$=$I_{\downarrow}$=$0$, which means that the system in this case
is an ordinary insulator without QSHE. We thus conclude that while the
previously studied topological winding number index $I_{c}$ (=$I_{\uparrow}%
$+$I_{\downarrow}$) \cite{Hatsugai,Wang,Hao} has been verified to describe the
IQH insulator, our present studied topological winding number index $I_{s}$
(=$I_{\uparrow}-I_{\downarrow}$) can be used to characterize the QSH insulator.

\begin{figure}[ptb]
\begin{center}
\includegraphics[width=0.6\linewidth]{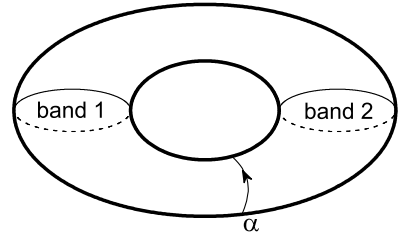}
\end{center}
\caption{The RS of the Bloch function corresponding to Fig. 8, in which the
edge states are sunk in the energy bands and can not form a loop around the
hole.}%
\end{figure}

Finally let us turn to consider the most complex case of $s_{z}$ being
nonconserved, i.e., the Rashba coupling $\lambda_{R}$, as well as the on-site
energy $\lambda_{\upsilon}$, is finite in the KM graphene model (\ref{Hr}).
After a straightforward derivation, we obtain the following couple eigenvalue
equations for $\lambda_{R}\neq0$:%
\begin{align}
(\epsilon\pm p_{2\pm})\Psi_{nA\pm}  &  =p_{1}\Psi_{nB\pm}+t_{1}\Psi
_{(n-1)B\pm}\nonumber\\
&  \pm p_{3}\left[  \Psi_{(n+1)A\pm}+\Psi_{(n-1)A\pm}\right] \nonumber\\
&  +ip_{\mp}\Psi_{nB\mp}-i\lambda_{R}\Psi_{(n-1)B\mp}, \label{14a}%
\end{align}%
\begin{align}
(\epsilon\mp p_{2\pm})\Psi_{nB\pm}  &  =p_{1}\Psi_{nA\pm}+t_{1}\Psi
_{(n+1)A\pm}\nonumber\\
&  \mp p_{3}\left[  \Psi_{(n+1)B\pm}+\Psi_{(n-1)B\pm}\right] \nonumber\\
&  -ip_{\pm}\Psi_{nA\mp}+i\lambda_{R}\Psi_{(n+1)A\mp}, \label{14b}%
\end{align}
where $p_{2\pm}$=$p_{2}\pm\lambda_{v}$, $p_{\pm}$=$2\lambda_{R}\cos
(\frac{\sqrt{3}}{2}ka\pm\frac{\pi}{3})$, and \textquotedblleft$\pm
$\textquotedblright\ in $\Psi_{nA(B)\pm}$ label the spin channels. Although
the Harper equation and the corresponding transform matrix are very difficult
to derive in this case, we can also distinguish the QSH phase from the
ordinary insulator phase by the topological winding index introduced in the
present paper directly from the energy spectrum. As an example, we
reinvestigate the QSH phase and the ordinary insulator phase with the Rashba
coupling being a finite value $\lambda_{R}$=$0.1t$. The other parameters are
chosen be the same as thosed used in Fig. 7. The calculated energy spectrum
(not shown here) turns to be essentially the same as that depicted in Fig. 7.
As a result, the corresponding RS of the Bloch wave functions is topologically
equivalent to Fig. 3. This means that in spite of finite Rashba term, the
system is still in the QSH phase with the topological index $I_{s}$=$2$.
Therefore, although in this case the spins are nonconserved (see Table I) and
the exact quantization of the spin Hall conductance is destroyed by the
spin-nonconserved perturbation, the QSH phase of the graphene system is still
topologically distinct from the ordinary insulating phase once provided that
the Rashba term does not change the system's topological properties. This
conclusion keeps consistent with Ref. \cite{Kane2}. If the Rashba spin-orbit
coupling $\lambda_{R}$ turns much strong, the concept of the \textquotedblleft
spin edge states\textquotedblright\ becomes faint and the winding number of
the spin edge states becomes meaningless. \begin{table}[th]
\caption{Comparison of the spin probability distribution ($\langle
s_{z}\rangle$) of the lowest-energy edge states on the lattice sites in the
cases $\lambda_{R}$=$0$ and $\lambda_{R}$=$0.1t$. The other parameters are set
as $\lambda_{\text{so}}$=$0.05t$ and $\lambda_{v}$=$0.1t$ and the momentum $k$
is kept as $k$=$0.99\pi$.}
%\label{specs}%
\begin{tabular}
[c]{cc|cc}\hline\hline
\multicolumn{2}{c|}{Case $\lambda_{R}$=$0$} & \multicolumn{2}{|c}{Case
$\lambda_{R}$=$0.1t$}\\\hline
lattice site index $n$ & $\langle s_{z}\rangle$ & lattice site index $n$ &
$\langle s_{z}\rangle$\\
$1$ & $\ 0.9998$ & $1$ & $0.9635$\\
$2$ & $\ 0.0001$ & $2$ & $-0.0274$\\
$3$ & $\ \ 0.0001\ $ & $3$ & $0.0001$\\
Total probability & $1.0000$ & Total probability & $0.9362$\\\hline
\end{tabular}
\end{table}

Before ending the present paper, we would like to stress that the topological
index $I_{s}$ describing the QSH phase can also be interpreted in terms of
Laughlin and Halperin's arguments \cite{Laughlin, Halperin}. In fact, in the
QSH system with topological integer $I_{s}$=$I_{\uparrow}-I_{\downarrow}$,
there are $I_{\uparrow}$ spin-up electrons transferred from one edge to the
other when a unit magnetic flux is adiabatically through a cylindrical system
by $I_{\uparrow}$ branches of gapless edge states. And at the same time there
are $|I_{\downarrow}|=I_{\uparrow}$ spin-down electrons to be transferred with
the opposite direction (see Fig. 10).\begin{figure}[ptb]
\begin{center}
\includegraphics[width=0.4\linewidth]{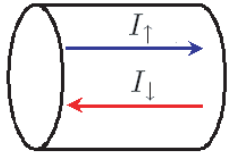}
\end{center}
\caption{(Color online) The Laughlin-Halperin diagram of the QSH system.}%
\end{figure}

In summary, by theoretically studying within the KM Hamiltonian a graphene
strip with zigzag edges, we have presented an alternative topological index
$I_{s}$ to characterize the QSH phase. The topological index $I_{s}$
describing QSH phases has been defined as the difference of the winding
numbers of the spin-resolved edge states crossing the holes of the
complex-energy RS. Based on this topological index, we have discussed
different phases by modulating different parameters in KM model, which agree
well with the previous studies in terms of the conventional $%
%TCIMACRO{\U{2124} }%
%BeginExpansion
\mathbb{Z}
%EndExpansion
_{2}$ topological invariance.

\begin{acknowledgments}
This work was supported by NSFC under Grants Nos. 10604010, 10534030, and
60776063, and the National Basic Research Program of China (973 Program) under
Grant No. 2009CB929103.
\end{acknowledgments}


\begin{thebibliography}{99}                                                                                               %


\bibitem {Klitzing}K. v. Klitzing, G. Dorda, and M. Peper, Phys. Rev. Lett.
\textbf{45}, 494 (1980).

\bibitem {Wen}X. G. Wen, Phys. Rev. B \textbf{40}, 7387 (1989).

\bibitem {TKNN}D. J. Thouless, M. Kohmoto, M. P. Nightingale, and M. den Nijs,
Phys. Rev. Lett. \textbf{49}, 405 (1982).

\bibitem {Chern}D. J. Thouless, \textit{Topological Quantum Numbers in
Nonrelatisvistic Physics} (World Scientific, Singapore, 1998).

\bibitem {Halperin}B. I. Halperin, Phys. Rev. B \textbf{25}, 2185 (1982).

\bibitem {Laughlin}R. B. Laughlin, Phys. Rev. B \textbf{23}, 5632 (1981).

\bibitem {Hatsugai}Y. Hatsugai, Phys. Rev. Lett. \textbf{71}, 3697 (1993);
Phys. Rev. B \textbf{48}, 11851 (1993).

\bibitem {Kane1}C. L. Kane and E. J. Mele, Phys. Rev. Lett. \textbf{95},
226801 (2005).

\bibitem {Kane2}C. L. Kane and E. J. Mele, Phys. Rev. Lett. \textbf{95},
146802 (2005).

\bibitem {Muk1}S. Murakami, N. Nagaosa, and S.C. Zhang, Science \textbf{301},
1348 (2003).

\bibitem {Sinova}J. Sinova, D. Culcer, Q. Niu, N. A. Sinitsyn, T. Jungwirth,
and A. H. MacDonald, Phys. Rev. Lett. \textbf{92}, 126603 (2004).

\bibitem {Mura2004}S. Murakami, N. Nagaosa, and S.-C. Zhang, Phys. Rev. Lett.
\textbf{93}, 156804 (2004).

\bibitem {Onoda2005}M. Onoda and N. Nagaosa, Phys. Rev. Lett. \textbf{95},
106601 (2005).

\bibitem {Bern20061}B. A. Bernevig and S.-C. Zhang, Phys. Rev. Lett.
\textbf{96}, 106802 (2006).

\bibitem {Qi2006}X.-L. Qi, Y.-S. Wu, and S.-C. Zhang, Phys. Rev. B
\textbf{74}, 085308 (2006).

\bibitem {Sheng2006}D. N. Sheng, Z. Y. Weng, L. Sheng, and F. D. M. Haldane,
Phys. Rev. Lett. \textbf{96}, 036808 (2006).

\bibitem {Fuk2007}T. Fukui and Y. Hatsugai, Phys. Rev. B \textbf{75},
121403(R) (2007).

\bibitem {Fu1}L. Fu and C. L. Kane, Phys. Rev. B \textbf{74}, 195312 (2006).

\bibitem {Fu2}L. Fu, C. L. Kane, and E. J. Mele, Phys. Rev. Lett. \textbf{98},
106803 (2007).

\bibitem {Fu3}L. Fu and C. L. Kane, Phys. Rev. B 76, 045302 (2007).

\bibitem {Mura2006}S. Murakami, Phys. Rev. Lett. \textbf{97}, 236805 (2006).

\bibitem {Onoda2007}M. Onoda, Y. Avishai, and N. Nagaosa, Phys. Rev. Lett.
\textbf{98}, 076802 (2007).

\bibitem {Wu2006}C. Wu, B. A. Bernevig, and S.-C. Zhang, Phys. Rev. Lett.
\textbf{96}, 106401 (2006).

\bibitem {Xu2006}C. Xu and J. E. Moore, Phys. Rev. B \textbf{73}, 045322 (2006).

\bibitem {Bern20062}B. A. Bernevig, T. L. Hughes, S.-C. Zhang, Science
\textbf{314}, 1757 (2006).

\bibitem {Moore2007}J. E. Moore and L. Balents, Phys. Rev. B \textbf{75},
121306 (R) (2007).

\bibitem {Haldane}F. D. M. Haldane, Phys. Rev. Lett. \textbf{61}, 2015 (1988).

\bibitem {Ohgushi}K. Ohgushi, S. Murakami, and N. Nagaosa, Phys. Rev. B
\textbf{62}, R6065 (2000).

\bibitem {Essin2007}A. M. Essin and J. E. Moore, Phys. Rev. B \textbf{76},
165307 (2007).

\bibitem {SCZhang}S. Murakami, N. Nagaosa, and S. C. Zhang, Science
\textbf{301}, 1348 (2003); Phys. Rev. B \textbf{69}, 235206 (2004).

\bibitem {Harper}P. G. Harper, Proc. Phys. Soc. London Sect. A \textbf{68},
874 (1955).

\bibitem {Yang}N. Byers and C. N. Yang, Phys. Rev. Lett. \textbf{7}, 46 (1961).

\bibitem {konig}M. K\"{o}nig, H. Buhmann, L. W. Molekamp, T. L. Hughes, C.-X.
Liu, X.-L. Qi, and S.-C. Zhang, arXiv/0801.0901.

\bibitem {creutz}M. Creutz and I. Horvath, Phys. Rev. D \textbf{50}, 2297
(1994); M. Creutz, Rev. Mod. Phys., \textbf{73}, 119 (2001).

\bibitem {Wang}Z. Wang and P. Zhang, Phys. Rev. B \textbf{77}, 125119 (2008).

\bibitem {Hao}N. Hao, P. Zhang, Z. Wang, W. Zhang, and Y. Wang, Phys. Rev. B
\textbf{78}, 075438 (2008).
\end{thebibliography}
\end{document}